\documentclass[letterpaper]{article}
\usepackage{spconf,amsmath,graphicx}
\usepackage[T1]{fontenc}
\usepackage[utf8]{inputenc}
\usepackage{microtype,booktabs,tabularx,array}
\usepackage{balance}
\usepackage[dvipsnames]{xcolor}
\usepackage{tikz}
\usetikzlibrary{arrows.meta,positioning,calc}
\usepackage[hidelinks]{hyperref}
\hypersetup{pdftitle={Voice Agents under Acoustic Stress: From Signal Degradation to Interaction and Action},pdfsubject={Version B: revised TRACE workflow and limitations},pdfkeywords={voice agents, acoustic robustness, spoken dialogue, tool use, evaluation}}
\definecolor{Ink}{HTML}{17324D}
\definecolor{Teal}{HTML}{116D70}
\definecolor{Rust}{HTML}{A74032}
\definecolor{Pale}{HTML}{F1F4F6}
\definecolor{Gray}{HTML}{485563}
\newcolumntype{Y}{>{\raggedright\arraybackslash}X}

\newcommand{\lead}[1]{\textbf{#1}}
\let\originalbibliography\thebibliography
\renewcommand{\thebibliography}[1]{\originalbibliography{#1}\setlength{\itemsep}{0pt}\setlength{\parsep}{0pt}}
\title{Voice Agents under Acoustic Stress:\\From Signal Degradation to Interaction and Action}
\name{
    \begin{tabular}{@{}c@{}}
    Amir Ivry$^{1}$ \qquad
    Kai-Wei Chang$^{2}$ \qquad
    Lin Zhang$^{3}$ \qquad
    Sharon Gannot$^{4}$ \qquad
    Carlos Busso$^{5}$
    \end{tabular}
}

\address{
    $^{1}$Technion -- Israel Institute of Technology \qquad
    $^{2}$Massachusetts Institute of Technology \\
    $^{3}$Independent Researcher \qquad
    $^{4}$Bar-Ilan University \qquad
    $^{5}$Carnegie Mellon University
}
\begin{document}
\ninept
\raggedbottom
\maketitle

\begin{abstract}
Voice agents must complete users' tasks despite noise, reverberation, and competing speech. Evaluating agents' robustness therefore requires following how acoustic conditions affect the conversation and the actions taken on the user's behalf. This overview examines what existing benchmarks reveal about agents' ability to complete tasks under acoustic stress and where further task-based evaluation is required. We then introduce TRACE, a practical workflow for designing, running, and interpreting evaluations of acoustic robustness in task-oriented human-agent interactions: the same agent attempts a specified task with an original recording and an acoustically stressed copy, and the resulting conversations are scored for task completion, wrong actions, recovery, and user effort. Finally, we explain how results from these evaluations can guide changes to an agent to prevent wrong actions and improve recovery.\footnote{Audio examples and code: \url{https://amir-ivry.github.io/trace/}}

\end{abstract}

\section{Introduction}
Suppose a voice assistant is asked to \textit{send a package to fifteen Oak Street}. Noise obscures the number, and the assistant hears \textit{send a package to \textbf{fifty} Oak Street}. If it places the delivery order immediately, it sends the package to the wrong address. If it first asks the user to repeat the number, it may obtain the correct address before placing the order, but the user must answer another question (Fig.~\ref{fig:consequence}). The outcome therefore depends on how the assistant responds to the noisy request. Assessing acoustic robustness requires following that response through to the task's outcome: does the assistant complete the task correctly, and how much effort does it require from the user? This overview explains how practitioners can design tests to answer that question for their own voice agents. 

Recent benchmarks have begun to address this broader view of robustness. VoiceBench examines how spoken requests and listening conditions affect answer quality \cite{chen2026voicebench}, while $\tau$-Voice and EVA-Bench follow task completion and conversational behavior in simulated interactions \cite{ray2026tauvoice,bogavelli2026evabench}. These broader tests matter because answering accurately does not guarantee that an agent knows when it needs more information. E.g., \emph{Pardon?} finds that models can answer questions successfully yet fail to request clarification when essential information is missing \cite{huang2026pardon} and LALM-as-a-Judge shows that automated detection of harmful spoken content depends heavily on the model used for judging and what its input modalities are \cite{ivry2026lalmasajudge}.

These studies provide complementary robustness tests, each with its own tasks, conditions, and scoring rules. For practitioners, the remaining challenge is to combine their lessons into a reproducible test of a particular application, with clear expectations for the agent's behavior and criteria for judging the outcome. We address this need by introducing TRACE (Fig.~\ref{fig:trace}), one possible practical workflow for evaluating acoustic robustness in task-oriented human-agent interactions. Our contributions are:
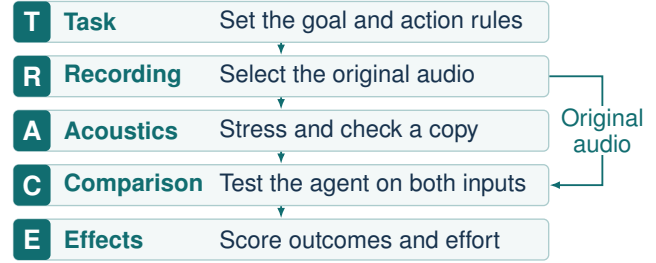
\begin{figure}[t]
\centering
% Editable TRACE flowchart with aligned step names and descriptions.
% The bypass retains the original audio for comparison.
\definecolor{TRACEAccent}{HTML}{116D70}
\begin{tikzpicture}[x=1cm,y=1cm,
  font=\sffamily\fontsize{9}{10.2}\selectfont,
  tracebox/.style={draw=TRACEAccent!35,fill=TRACEAccent!5,
    rounded corners=2pt,line width=.5pt,
    minimum width=7.08cm,minimum height=.54cm,inner sep=0pt},
  flow/.style={-{Latex[length=1.4mm]},draw=TRACEAccent,line width=.7pt}]
\foreach \id/\y/\letter/\stepname/\description in {
  t/0/T/Task/Set the goal and action rules,
  r/-.72/R/Recording/Select the original audio,
  a/-1.44/A/Acoustics/Stress and check a copy,
  c/-2.16/C/Comparison/Test the agent on both inputs,
  e/-2.88/E/Effects/Score outcomes and effort}{
  \node[tracebox] (\id) at (3.54,\y) {};
  \fill[TRACEAccent,rounded corners=2pt]
    (0,\y-.27) rectangle (.49,\y+.27);
  \node[text=white,font=\sffamily\bfseries\fontsize{10}{11}\selectfont]
    at (.245,\y) {\letter};
  \node[anchor=west,inner sep=0pt,text=TRACEAccent,font=\sffamily\bfseries\fontsize{9}{10.2}\selectfont]
    at (.66,\y) {\stepname};
  \node[anchor=west,inner sep=0pt,text=Ink]
    at (2.72,\y) {\description};
}
\draw[flow] (t.south) -- (r.north);
\draw[flow] (r.south) -- (a.north);
\draw[flow] (a.south) -- (c.north);
\draw[flow] (c.south) -- (e.north);
\draw[flow] (r.east) -- (7.79,-.72) -- (7.79,-2.16) -- (c.east);
\node[fill=white,align=center,inner sep=1pt,text=TRACEAccent]
  at (7.79,-1.44) {Original\\audio};
\end{tikzpicture}
\caption{\textbf{The TRACE workflow.} The original audio is retained while a copy is stressed acoustically. Both versions are tested with the same agent, and are compared by task outcomes, recovery, and user effort.}
\label{fig:trace}
\end{figure}

\begin{itemize}\setlength{\itemsep}{0pt}\setlength{\parsep}{0pt}
\item \textbf{A guide to what needs testing.}
We explain how acoustic stress can affect an agent's decisions and compare which behaviors existing benchmarks test. This framework helps practitioners identify what remains to be evaluated for their application (Sec.~\ref{sec:evidence}).

\item \textbf{TRACE: a workflow for carrying out those tests.}
We provide steps for designing, running, and interpreting acoustic robustness tests, and show how to assess task completion, wrong actions, recovery, and user effort (Sec.~\ref{sec:protocol}).

\item \textbf{Guidance for using the results.}
We explain how failures found during TRACE can guide changes to the agent and how to check whether those changes help (Sec.~\ref{sec:mitigation}). 
\end{itemize}
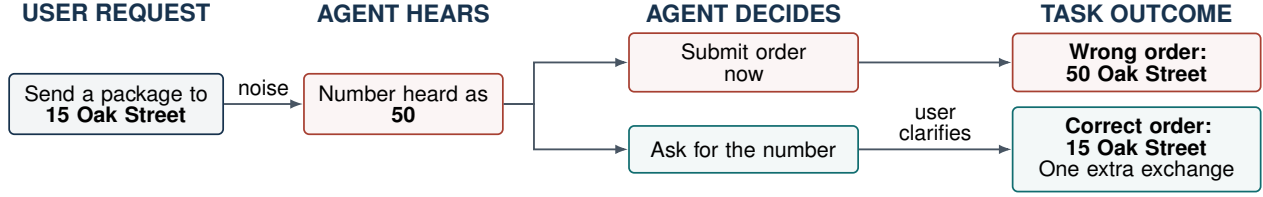
\begin{figure*}[t]
\centering
% Editable vector artwork; 9 pt labels at native size.
\begin{tikzpicture}[x=1cm,y=1cm,font=\sffamily\fontsize{8}{8.5}\selectfont,
 box/.style={rounded corners=2pt,draw=#1,fill=#1!5,line width=.65pt,align=center,inner sep=4pt},
 arrow/.style={-{Latex[length=1.7mm]},line width=.7pt,draw=Gray}]
\node[text=Ink,font=\sffamily\bfseries\fontsize{9}{10.4}\selectfont] at (1.5,1.95) {USER REQUEST};
\node[text=Ink,font=\sffamily\bfseries\fontsize{9}{10.4}\selectfont] at (5.3,1.95) {AGENT HEARS};
\node[text=Ink,font=\sffamily\bfseries\fontsize{9}{10.4}\selectfont] at (9.8,1.95) {AGENT DECIDES};
\node[text=Ink,font=\sffamily\bfseries\fontsize{9}{10.4}\selectfont] at (15,1.95) {TASK OUTCOME};
\node[box=Ink,text width=2.55cm,minimum height=.8cm] (goal) at (1.5,.75) {Send a package to\\\textbf{15 Oak Street}};
\node[box=Rust,text width=2.35cm,minimum height=.8cm] (heard) at (5.3,.75) {Number heard as\\\textbf{50}};
\node[box=Rust,text width=2.75cm,minimum height=.62cm] (act) at (9.8,1.3) {Submit order\\now};
\node[box=Teal,text width=2.75cm,minimum height=.62cm] (ask) at (9.8,.15) {Ask for the number};
\node[box=Rust,text width=3.0cm,minimum height=.62cm] (bad) at (15,1.3) {\textbf{Wrong order: 50 Oak Street}};
\node[box=Teal,text width=3.0cm,minimum height=.62cm] (good) at (15,.15) {\textbf{Correct order: 15 Oak Street}\\One extra exchange};
\draw[arrow] (goal) -- node[above]{noise} (heard);
\draw[arrow] (heard.east) -- ++(.4,0) |- (act.west);
\draw[arrow] (heard.east) -- ++(.4,0) |- (ask.west);
\draw[arrow] (act) -- (bad);
\draw[arrow] (ask) -- node[above,align=center]{user\\clarifies} (good);
\end{tikzpicture}
\caption{\textbf{How one misheard number due to acoustic stress affects the user-agent conversation.} In this hypothetical example, the assistant hears \textit{fifty} instead of \textit{fifteen} in the delivery address. If it places the order immediately, the package goes to the wrong address. If it first asks the user to repeat the number and understands the reply, it places the order correctly, but the user must answer an extra question.}
\label{fig:consequence}
\end{figure*}

\section{How acoustic stress affects agent behavior}
\label{sec:failures}
The request to send a package to fifteen Oak Street in Fig.~\ref{fig:consequence} illustrates how a misheard number can lead to a wrong action. Acoustic stress can disrupt three related decisions: what does the agent understand? whom does it follow? and when does it act?

\noindent\lead{Content: what does the agent understand?}
Noise and reverberation can obscure the sounds that distinguish one word from another. The agent may then interpret ``fifteen'' as ``fifty'' and use the wrong address when placing the order. The problem can extend beyond substituting similar words: WildASR documents transcriptions containing words that were never spoken when recordings are clipped or cut short \cite{tay2026wildasr}. An agent relying on such a transcription may therefore act on information the user never supplied.

\noindent\lead{Source: whom does the agent follow?}
Competing speech can cause an agent to attribute a bystander's words to the user. Lee et al.~\cite{lee2026still} show that voice assistants can mistake another speaker's interjection for a correction by the original user. In the package example, a bystander's ``fifty'' could then replace the user's ``fifteen,'' even if the agent hears both numbers correctly. The resulting order would follow the wrong person's instruction.

\noindent\lead{Timing: when does the agent act?}
Noise or overlapping speech can obscure the start of a user correction, making an unfinished request appear complete. If the user says ``fifty--sorry, fifteen,'' the agent may place the order before hearing the corrected number. Full-Duplex-Bench-v3 illustrates the consequence of premature action: an early tool call can retain an outdated destination despite a subsequent correction \cite{lin2026fdbv3}. Understanding the correction later does not prevent the initial wrong action. The key challenge, therefore, is deciding when to act and when to wait for possible corrections before committing to an irreversible tool call.
% These failures can combine when, for example, a bystander supplies the wrong number while masking the user's correction.

% \noindent\lead{Recovery: can the user understand the clarification?}
% Clarification depends on the user understanding the agent's spoken question. Background noise may mask the distinction in ``fifteen or fifty?'', causing the user to ask for repetition instead of resolving the address. The agent has chosen an appropriate question, but the exchange adds effort without correcting the misunderstanding. Checking only the text of the agent's response would miss this failure because it does not establish what the user actually heard.

\section{What existing benchmarks tell us}
\label{sec:evidence}
\begin{table*}[t]
\centering
\caption{\textbf{What existing benchmarks test, from speech quality to task completion.} The final column identifies what their scores do not verify, such as whether a correct tool call actually completes the task.}
\label{tab:evidence}
\begingroup
\renewcommand{\arraystretch}{1}
\begin{tabularx}{\textwidth}{@{}>{\raggedright\arraybackslash}p{.25\textwidth}>{\raggedright\arraybackslash}p{.39\textwidth}Y@{}}
\toprule
\textbf{What is tested}
% \textbf{Measured outcome}
& \textbf{Benchmarks that test it}
& \textbf{What remains untested}\\
\midrule
Audio quality, intelligibility, recognition, or speaker attribution
& REVERB; DNS; CHiME-7; NOTSOFAR; AEC; TTS with noisy references \cite{kinoshita2013reverb,reddy2020dns,cornell2023chime7,vinnikov24_interspeech,cutler2024aec,wang2024ttsnoise}
& Whether these abilities lead to successful agent tasks.\\
\midrule[0.25pt]

Answer quality, including use of earlier dialogue information
& VoiceBench; VoxEval; Audio MultiChallenge \cite{chen2026voicebench,cui2025voxeval,gosai2026audiomultichallenge}
& Whether the agent executes the correct action and reaches the intended task state.\\
\midrule[0.25pt]

Correct tools and arguments; rejection of side conversations % in WearVox
& WearVox; Audio2Tool; Text to Voice \cite{lin2026wearvox,pahwa2026audio2tool,laskar2026texttovoice}
& Whether the proposed calls execute successfully and complete the task.\\
\midrule[0.25pt]

Turn-taking and response timing; timed tool calls % in v3
& Full-Duplex-Bench series \cite{lin2025fullv1,lin2026fullv1.5,lin2026fullv2,lin2026fdbv3}
& Whether action timing remains reliable with real service delays and failures.\\
\midrule[0.25pt]

Clarification requests or transcript correction
& \emph{Pardon?}; Interactive ASR \cite{huang2026pardon,wang2026interactive}
& Whether the correction leads to successful task recovery.\\
\midrule[0.25pt]

Task completion or correct final database values
& $\tau$-Voice; EVA-Bench; $\tau$-Elicit \cite{ray2026tauvoice,bogavelli2026evabench,ray2026tauelicit}
& Whether results generalize beyond the tested tasks, acoustic conditions, and interactions.\\
\bottomrule
\end{tabularx}
\endgroup
\end{table*}
Acoustic stress can make an agent misunderstand a request, follow the wrong speaker, or act before a correction ends. Tab.\ref{tab:evidence} includes both speech-processing benchmarks, which help diagnose problems in audio quality or recognition, and voice-agent benchmarks, which assess responses and actions. Both are relevant because identifying a speech-processing problem does not establish its consequence for the task. We draw three conclusions from these tests and benchmarks: errors observed under noisy conditions are not necessarily caused by noise, task completion alone does not characterize the entire interaction, and requesting clarification does not necessarily indicate successful task recovery. These conclusions are detailed below.

\noindent\lead{Task accuracy and sensitivity to noise are different.}
To estimate whether acoustic stress adds errors, an evaluation must compare matched versions of the same task with and without the stressor. Laskar et al. make this comparison for tool calling by converting written tool-use requests into speech and adding noise~\cite{laskar2026texttovoice}. Their results illustrate that tool-calling performance changes only modestly across the tested noise levels. Thus, performance with noisy audio measures task accuracy under that condition, whereas the difference from the matched clean condition estimates sensitivity to noise.

\noindent\lead{Task completion does not describe the whole interaction.}
$\tau$-Voice measures task completion alongside how agents handle simulated conversations \cite{ray2026tauvoice}. Under its realistic test conditions, the agent with the highest overall task completion rate also interrupts the user most often. Completing more tasks therefore does not establish that an agent handles the conversation better: interruptions must be measured separately from the final task outcome.

\noindent\lead{Clarification and task recovery are different outcomes.}
Recovery involves both obtaining a correction and using it to complete the task, but evaluations may examine only part of this process. \emph{Pardon?} tests whether missing essential information prompts a clarification request \cite{huang2026pardon}. Beyond requesting help, Interactive ASR tests whether spoken feedback corrects a transcript \cite{wang2026interactive}, while $\tau$-Elicit follows value collection through conversation to the final database submission \cite{ray2026tauelicit}.
For the example in Fig.~\ref{fig:consequence}, neither a clarification request nor a corrected transcript establishes that the right order was placed or that no wrong order preceded it.

A robustness test should show whether acoustic stress leads to additional wrong actions, fewer completed tasks, or excess work for the user. This evaluation requires comparing the same agent on the same tasks with and without added stress and checking what it does throughout each conversation, including mistakes it later corrects.
TRACE organizes these checks into a practical workflow for testing a voice agent on the tasks it is intended to perform.

\section{TRACE: a workflow for testing voice agents}
\label{sec:protocol}
TRACE tests how acoustic stress affects task completion through five steps: \emph{Task, Recording, Acoustics, Comparison, and Effects}. Fig.~\ref{fig:trace} summarizes the workflow and Tab.\ref{tab:outcomes} develops the delivery example in Fig.~\ref{fig:consequence} into a test with explicit scoring rules.

The example compares the original request for 15 Oak Street with a copy in which ``fifteen'' is replaced by silence. It tests whether the agent asks for the missing number and uses the reply before placing the order. Unlike adding noise, which may leave the number understandable, this deliberate removal creates a need for clarification.
% \input{figures/trace}
% Self-contained table. All scenarios and aggregate counts are hypothetical.
\begin{table*}[t]
\centering
\caption{\textbf{A worked TRACE example for delivery to 15 Oak Street.} The first part defines the task, speech inputs, and comparison setup. The second part shows how five possible human-agent conversations are scored. The third part illustrates how these scores are summarized across ten independent recordings, each tested with original audio and with the street number removed. All outcomes and counts are hypothetical.
% Recovery rates give the percentage of initiated attempts that complete recovery. Extra user turns are counted after the initial instruction; the third part reports their average across conversations in each audio condition.
}
\label{tab:outcomes}
\begingroup
\setlength{\tabcolsep}{3pt}
\setlength{\aboverulesep}{1.5pt}
\setlength{\belowrulesep}{1.5pt}
\renewcommand{\arraystretch}{1.05}
\newcommand{\TRACErate}[2]{\begin{tabular}[t]{@{}c@{}}#1\\(#2\%)\end{tabular}}
\begin{tabularx}{\textwidth}{@{}>{\raggedright\arraybackslash}p{.12\textwidth}Y@{}}
\toprule
\textbf{T: Task} & One active order for \textbf{15 Oak Street} within time and turn limits. Prohibited: wrong orders, ordering before obtaining address.\\
\textbf{R: Recording} & ``Please send a package to \textbf{fifteen Oak Street}.'' The address' start and end positions are marked. No other context reveals it.\\
\textbf{A: Acoustics} & The original audio is retained; ``fifteen'' is replaced by silence in a copy. Remaining audio and context are checked for clues.\\
\textbf{C: Comparison} & Both runs start with a fresh conversation using the same agent settings and user-response rules. A user's willingness to clarify stays fixed within each pair. Replies from both ends receive no added acoustic stress.\\
\bottomrule
\end{tabularx}
\par\vspace{2pt}
\begin{tabularx}{\textwidth}{@{}c>{\raggedright\arraybackslash}p{.083\textwidth}Ycc>{\centering\arraybackslash}p{.087\textwidth}cc@{}}
\toprule
\textbf{Scenario} & \shortstack[l]{\textbf{Input}\\\textbf{audio}} & \textbf{Agent behavior and user response}
& \shortstack{\textbf{Goal}\\\textbf{attained}}
& \textbf{Violation}
& \textbf{Recovery}
& \shortstack{\textbf{Task}\\\textbf{success}}
& \shortstack{\textbf{Extra user}\\\textbf{turns}}\\
\midrule
1 & Original & Places the correct order without clarification.
& Yes & No & No attempt & Yes & 0\\
2 & Number removed & Asks for the number; the user provides it; agent places the correct order.
& Yes & No & Completed & Yes & 1\\
3 & Original & Places a wrong order; the user corrects it; agent cancels and replaces the order correctly.
& Yes & Yes & Completed & No & 1\\
4 & Number removed & Asks for the number; the user refuses to continue; no order is placed.
& No & No & Incomplete & No & 1\\
5 & Number removed & Guesses the number correctly and orders without clarification.
& Yes & Yes & No attempt & No & 0\\
\midrule
\multicolumn{8}{@{}l@{}}{\textbf{Metrics across ten paired recordings: ten conversations per condition}}\\
\midrule
 & Original & Scenario~1:~9 conversations; scenario~3:~1.
& \TRACErate{10/10}{100} & \TRACErate{1/10}{10}
& \TRACErate{1/1}{100} & \TRACErate{9/10}{90} & 0.1\\
 & Number removed & Scenario~2:~6 conversations; scenario~4:~2; scenario~5:~2.
& \TRACErate{8/10}{80} & \TRACErate{2/10}{20}
& \TRACErate{6/8}{75} & \TRACErate{6/10}{60} & 0.8\\
\bottomrule
\end{tabularx}
\endgroup
\end{table*}

\subsection{T---Task: The goal and rules for completing the goal}\label{sec:T}
A test first defines correct behavior and permitted actions to achieve the user's goal~\cite{walker1997paradise}. 
In Tab.\ref{tab:outcomes}, the goal is exactly one active order for 15 Oak Street, using a number provided by the user. 
Accordingly, the agent may neither guess the removed number, even if correct, nor place a wrong order, even if it is later canceled.
% Guessing the removed number is prohibited even if correct, as is placing a wrong order that is later cancelled. 
Time and user-turn constraints determine when an unfinished attempt ends. Jointly, these rules define success before testing begins.

\subsection{R---Recording: The speech that the agent receives}\label{sec:R}
The recording step establishes the original speech input before additional stress is introduced. Recordings are selected to represent the application's spoken instructions and users, with existing room and microphone effects documented as the starting acoustic conditions \cite{cornell2023chime7}.
In our example, the recording contains the user's spoken instruction, ``Please send a package to fifteen Oak Street.'' The start and end of the word ``fifteen'' are marked so that it can later be removed without changing the rest of the recording. The correct address is retained for scoring but is not provided to the agent.

Preserving the original audio and marking the relevant word provides a basis for applying a controlled change to a localized audio segment and testing how it affects the agent's responses and actions.
% Preserving the original audio and marking the relevant word provides a basis for applying a controlled acoustic change and testing how it affects the agent's responses and actions.

\subsection{A---Acoustics: The acoustic modification to the original speech and its effect on the available information}
\label{sec:validity}
The recording in \ref{sec:R} is the \emph{reference input}, whereas a stressed copy contains the chosen acoustic modification. 
In Tab.\ref{tab:outcomes}, the original samples for ``fifteen'' are replaced with zeros to simulate a brief, packet-loss-like audio dropout, preserving the request's duration and the remaining context.
% In Tab.\ref{tab:outcomes}, the original samples for ``fifteen'' are replaced with zeros to imitate transient noise, preserving the request's duration and the rest context.

The expected agent behavior depends on what the modified input supports \cite{ribeiro2020checklist}. The target stays unchanged when sufficient information remains (\emph{answer-preserving stress}) \cite{ivry2026taskawareanswerpreservationaudio}, whereas missing information requires a permitted way to recover it, such as clarification (\emph{missing evidence}) \cite{huang2026pardon}. A change that alters the answer itself (\emph{changed scene}) requires a revised target; e.g., adding a speaker to a speaker-counting task is scored against the new count, separately from tests requiring an unchanged answer.

The changed audio and all other information available to the agent are examined to determine what information remains available. In the delivery example, clarification is required only if the address' number cannot be determined from either the remaining recording or the available context.

\subsection{C---Comparison: Testing original and stressed recordings}
\label{sec:execution}
The comparison holds the agent and task fixed while changing the audio \cite{laskar2026texttovoice}. Each delivery run starts with a fresh conversation, using identical model settings. The original recording of the user's instruction and its stressed copy form a matched input pair.

For each recording of the delivery instruction, the same agent performs the task in two separate runs. Both start with a fresh conversation and no delivery order. One run receives the original audio, including ``fifteen'' and the other receives a copy with that word replaced by silence. The task and agent settings are otherwise identical.

If the agent asks for clarification, the conversation continues through the user's reply \cite{bohus2005recovery}. 
The simulated user's response rule stays fixed within each pair: cooperative users supply ``fifteen'' when asked, whereas in abandonment cases the interaction ends without a reply.
% The simulated user's response rule stays fixed within each pair: cooperative users supply ``fifteen'' when asked, whereas users in the abandonment cases refuse to continue. 
These replies receive no added stress. The comparison therefore assesses how removing the number changes the agent's actions, including whether it obtains the missing information through clarification.

\subsection{E---Effects: Task completion, recovery, and user effort}
\label{sec:scoring}\label{sec:comparison}
TRACE measures how a chosen acoustic condition affects task completion and user effort across multiple test cases. Tab.\ref{tab:outcomes} first uses five hypothetical delivery scenarios to explain how individual conversations are scored. It then illustrates how these scores become overall performance measures. For this purpose, the table assumes ten recordings of the delivery instruction, each tested with original audio and with the street number removed. Counting the successful conversations and averaging the extra user turns gives the success rate and user effort in each condition. Comparing these measures illustrates the effect of removing the number.

\noindent\lead{Task completion and violations.}
TRACE's scoring distinguishes reaching the requested result from completing the task without prohibited actions. We use \emph{goal attainment} for reaching that result within the task's time and turn limits, and \emph{violation} for any prohibited action. \emph{Task success} requires goal attainment without violations.

In scenario~3 in Tab.\ref{tab:outcomes}, the agent cancels an incorrect delivery order and replaces it with one for the correct address. This scenario attains the goal, but the earlier wrong-address submission prevents task success. Scenario~5 in Tab.\ref{tab:outcomes} also fails because guessing the missing street number violates the requirement to obtain it from the user, even when the guess is correct.

Scoring checks both the final task result and the actions taken throughout the conversation. The percentages of conversations that attain the goal, contain a violation, and achieve task success are calculated separately for the original recordings and their stressed copies.

\noindent\lead{Recovery.}
A \emph{recovery attempt} begins with clarification by the agent or correction by the user. It completes when the supplied information is used to continue the task. Scenarios~2 and~3 in Tab.\ref{tab:outcomes} complete recovery, but scenario~3 retains its violation. Scenario~4 ends without the missing information.
The \emph{recovery-completion rate} counts completed attempts among all attempts: six of eight, or 75\%, in the table's stressed condition. Trials without an attempt are excluded. Successful trials are also distinguished by whether recovery occurred. Because agents may encounter different errors, this rate alone cannot rank recovery strategies.

\noindent\lead{User effort and delay.}
Effort includes extra user turns and repeated information, including failed recovery \cite{walker1997paradise}. The table counts turns after the initial instruction: scenarios~2-4 each add one, including the refusal to continue. Delay is measured between stated events, such as the end of user speech and the start of the agent's reply. Premature responses are reported separately so interruptions cannot appear as faster service.

\noindent\lead{The effect of acoustic stress.}
Each recording is tested under both conditions with the same task and user-response rules. In Tab.\ref{tab:outcomes}, success falls from nine of ten conversations (90\%) to six (60\%): a 30-percentage-point decline. Violations rise from 10\% to 20\%, and mean extra turns from 0.1 to 0.8. These hypothetical counts illustrate the calculation, not measured performance. Actual comparisons report uncertainty and include timed-out and abandoned trials in the scores.

\balance

\section{Using TRACE results to improve the agent}
\label{sec:mitigation}
TRACE can help improve an agent by showing what goes wrong and whether a change helps. The workflow is applied before and after the change using the same test cases. Two examples illustrate this process for the delivery task in Fig.~\ref{fig:consequence}.

\noindent\lead{Improving how the agent hears the street number.}
Suppose TRACE reveals that the agent uses the wrong street number even though the number remains audible. One possible change is to add speech enhancement and target speaker extraction (TSE) before the audio reaches the agent. TRACE is then repeated with enhancement and TSE both enabled and disabled, using the same recordings under both original and stressed audio conditions. The resulting task-success rates and response delays show whether this front-end processing helps the agent complete the task and the delay it brings.

\noindent\lead{Improving how the agent obtains a missing number.}
Suppose TRACE instead reveals that asking for the entire instruction again often leaves the number unresolved. The agent is revised to ask directly, ``What is the street number?'' \cite{bohus2005recovery}. Applying TRACE to both clarification policies, with the same tasks, acoustic conditions, and rules for user replies, shows whether the targeted question increases task success or reduces additional user turns. Recovery is assessed through the agent's use of the answer to complete the delivery task.

For either change, TRACE is also applied to cases that were not used to develop it. This tests whether the improvement extends to new inputs and whether tasks completed correctly before still succeed.

\section{Limitations}
\label{sec:limitations}
TRACE's guidance depends on how well the selected robustness tests represent the intended application. For example, testing noise only in the initial instruction may miss failures that occur when clarification replies are also noisy. Improvements identified through TRACE therefore support conclusions about the tested tasks and conditions. Their usefulness in other settings requires further assessment.

\section{Conclusion}
\label{sec:conclusion}
TRACE provides one practical workflow for evaluating acoustic robustness in task-oriented human-agent interactions through what an agent accomplishes, the errors it makes, and the effort required from its user.
The next step is to test TRACE empirically across different voice agents, tasks, and acoustic conditions. These studies should compare TRACE with existing evaluation practices to determine whether it reveals additional failures and helps practitioners improve task success, while measuring any change in user effort.

\clearpage
\section{Acknowledgments}
\noindent\textbf{Funding:} No funding was received for conducting this study. The authors have no relevant financial or nonfinancial interests to disclose.

\vspace{1pt}\noindent\textbf{Ethics:} No human-subject data were collected as part of this study.

\vspace{1pt}\noindent\textbf{Use of AI:}  ChatGPT (OpenAI) was used for language editing. All AI-assisted outputs were reviewed and validated by the authors, who take full responsibility for the paper's content.
\balance
\bibliographystyle{IEEEbib}
\bibliography{references}
\end{document}